\documentclass{appolb}
\usepackage{graphicx}
\usepackage[hidelinks]{hyperref}

\begin{document}
\title{Charm production: constraints to transport models and charm diffusion coefficient with ALICE%
\thanks{Presented at Quark Matter 2022. The XXIXth International Conference on Ultra-relativistic Nucleus-Nucleus Collisions. In Krakow, Poland}%
}

\author{Luuk Vermunt
\address{University of Heidelberg (Germany)}
\\[3mm]
On behalf of the ALICE Collaboration
}

\maketitle

\begin{abstract}
In this contribution, the nuclear modification factor and azimuthal anisotropy of prompt charm mesons and baryons in Pb--Pb collisions at $\sqrt{s_\mathrm{NN}}=5.02$~TeV by the ALICE Collaboration are presented. Heavy quarks are a very suitable probe to investigate the quark--gluon plasma (QGP) produced in heavy-ion collisions, since they are mainly produced in hard-scattering processes and hence in shorter timescales compared to the QGP. Measurements of charm-hadron production in nucleus--nucleus collisions are therefore useful to study the properties of the in-medium charm-quark energy loss via the comparison with theoretical models. Models describing the heavy-flavour transport and energy loss in a hydrodynamically expanding QGP require also a precise modelling of the in-medium hadronisation of heavy quarks, which is investigated via the measurement of prompt $\mathrm{D_s^+}$ mesons and $\Lambda_\mathrm{c}^{+}$ baryons.
\end{abstract}
  
\section{Introduction}
In ultra-relativistic heavy-ion collisions, a phase transition of nuclear matter to a colour-deconfined medium is predicted, the so-called quark--gluon plasma (QGP)~\cite{Busza:2018rrf}. Heavy quarks (charm and beauty) are predominantly produced in the early stages of such collisions via hard-scattering processes. Due to the very short time scales characterising heavy-quark production, which are shorter than the QGP formation time (approximate $0.1$ and $0.03$~fm$/c$ for charm and beauty quarks~\cite{Andronic:2015wma}, and between $0.3$ and $1.5$~fm$/c$ for the QGP~\cite{Liu:2012ax}), heavy quarks experience the full evolution of the medium. Once produced, these quarks traverse the medium and interact via inelastic and elastic processes with its constituents. They are therefore an effective probe to study several aspects of the medium, like the properties of the energy loss mechanisms, the relevance of quark recombination in the hadronisation of the QGP, and the initial conditions of the system.

In this contribution, the most recent measurements of open-charm meson and baryon production by the ALICE Collaboration in the latest LHC Pb--Pb run at $\sqrt{s_{\rm NN}} = 5.02$~TeV (from 2018) are discussed~\cite{ALICE:2021rxa,ALICE:2020iug,ALICE:2021kfc,ALICE:2021bib}. In particular, i) the nuclear modification factor ($R_{\rm AA}$), defined as the ratio of the production yield in Pb--Pb collisions and the cross section in pp collisions scaled by the average nuclear overlap function $T_{\rm AA}$, ii) the elliptic flow ($v_2$), which is the second harmonic Fourier coefficient of the azimuthal anisotropies in the production of heavy-flavour hadrons, and iii) the baryon-to-meson production yield ratios will be presented. The first two observables are especially sensitive to the interactions of the heavy quarks with the QGP, while the hadron relative abundances provide information regarding possible modifications of the hadronisation mechanisms in presence of such a deconfined QCD medium~\cite{Greco:2003vf}.

Open charm hadrons are measured by ALICE at midrapidity ($|y|<0.5$) via the decay channels ${\rm D}^0  \to {\rm K}^- \pi^+$, ${\rm D}^+ \to {\rm K}^- \pi^+ \pi^+$, ${\rm D}^{*+} \to {\rm D}^0 \pi^+$, $\rm D_{s}^+ \to \phi \pi^+ \to K^{+}K^{-} \pi^+$, ${\rm \Lambda_{c}^+ \to pK^{0}_{s} \to p\pi^+\pi^-}$, ${\rm \Lambda_{c}^{+}\to p K^-\pi^+}$, and their charge conjugates. Topological and particle-identification selections are used to enhance the signal-to-background ratio, either via so-called rectangular selections or using machine-learning algorithms~\cite{Chen:2016btl}. The raw charm-hadron yields are extracted via invariant-mass analyses and corrected for the reconstruction and selection efficiency (estimated using Monte Carlo simulations) and the prompt fraction (based on a theory-driven method~\cite{ALICE:2021rxa,ALICE:2021bib}). The measurements of the $\rm D$-meson elliptic flow are performed with the scalar-product method~\cite{Voloshin:2008dg}.

\section{Results}

\begin{figure}[tb!]
  \centerline{%
     \includegraphics[trim={0 0 10cm 0},clip,width=0.5\textwidth]{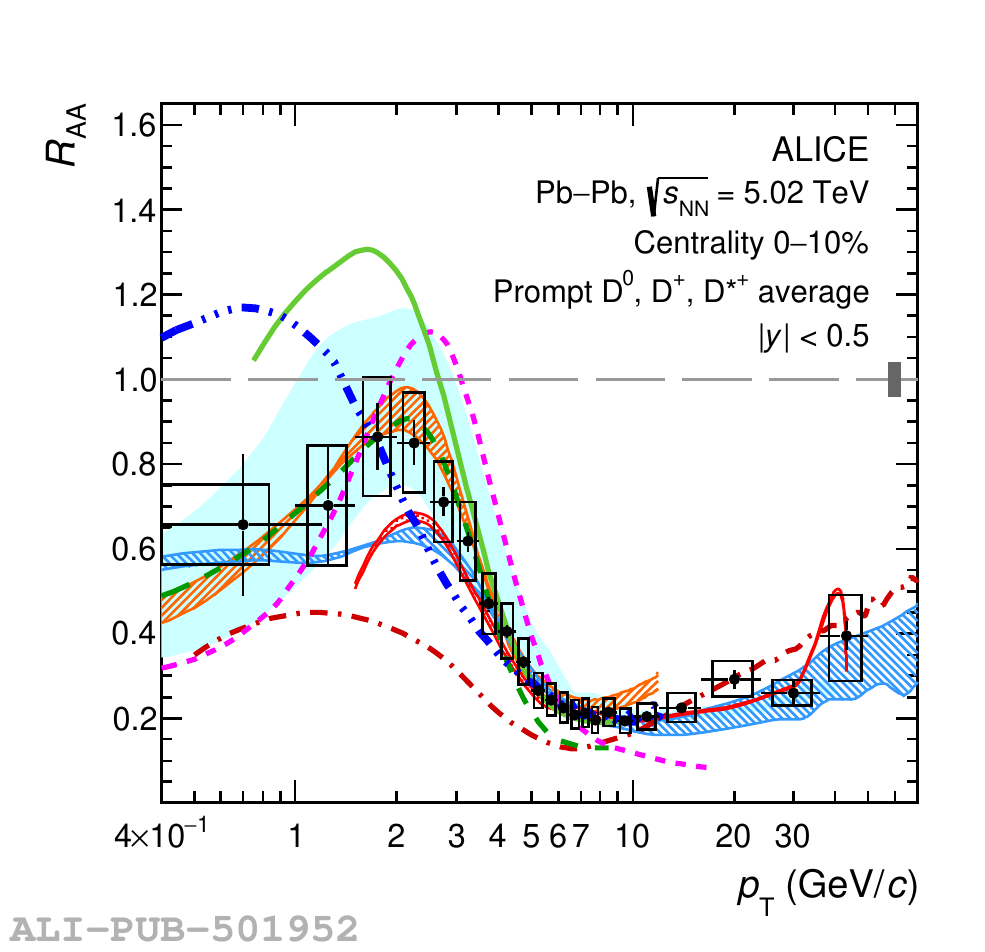}%
     \includegraphics[trim={10cm 0 0 0},clip,width=0.5\textwidth]{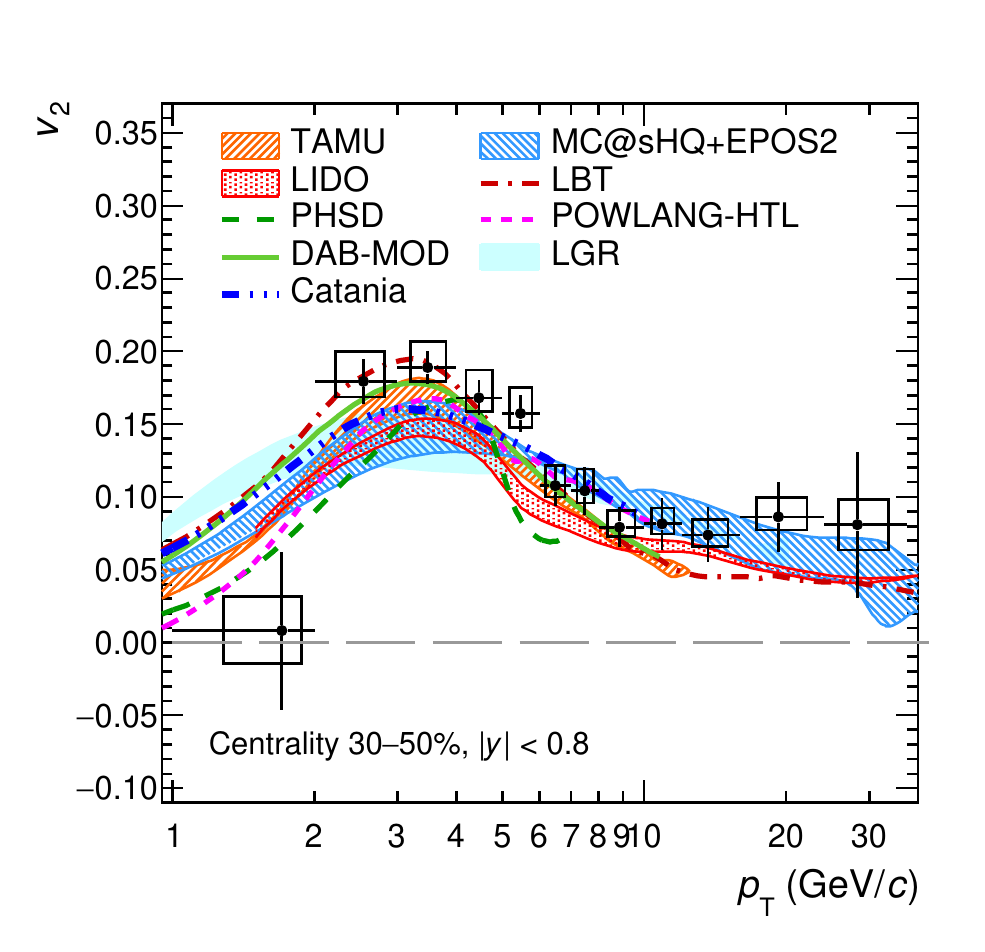}
  }
  \caption{The average $R_{\rm AA}$ (left) and $v_2$ (right) of prompt $\rm D^0$, $\rm D^+$, and $\rm D^{*+}$ mesons in, respectively, the 0--10\% and 30--50\% centrality classes~\cite{ALICE:2021rxa} compared with predictions of charm-quark transport models~\cite{Dmesonmodels}.}
  \label{fig:DRAAv2Theory}
\end{figure}

In Fig.~\ref{fig:DRAAv2Theory}, the nuclear modification factor in central Pb--Pb collisions and the elliptic flow in mid-central Pb--Pb collisions for prompt non-strange ${\rm D}$ mesons (average of $\rm D^0$, $\rm D^+$, and $\rm D^{*+}$) are shown~\cite{ALICE:2021rxa,ALICE:2020iug}. The $R_{\rm AA}$, measured for the first time down to $p_{\rm T}=0$, shows a suppression of a factor 5 with respect to the binary-scaled pp reference, with a minimum value at $p_{\rm T} \approx 7$~GeV$/c$. The measured $v_2$ is found to be significantly larger than zero in the $2 < p_{\rm T} < 24$~GeV$/c$ interval. The measurements are compared to various predictions from models implementing charm-quark transport in a hydrodynamically expanding medium~\cite{Dmesonmodels}. The main differences between these models are i) the use of solely collisional or collisional and radiative interaction processes, ii) the inclusion of initial-state effects by using nuclear parton distribution functions, and iii) the way hadronisation via quark recombination (in addition to charm-quark fragmentation) is implemented. Most of the models capture the magnitude and $p_{\rm T}$ trend of the $R_{\rm AA}$ for $p_{\rm T} > 6$~GeV$/c$, while there are significant deviations at lower $p_{\rm T}$. The models describe reasonably well the $v_2$, even though they tend to slightly underestimate the data in the $2 < p_{\rm T} < 6$~GeV$/c$ interval. By considering the few models that are in fair agreement with both the nuclear modification factor and the elliptic flow (based on a $\chi^2 / {\rm ndf} < 5$ and $< 2$ requirement, respectively), the heavy-quark spatial diffusion coefficient, $D_s$, was estimated to be in the range $1.5 < 2 \pi D_s T_{\rm c} < 4.5$ at the pseudocritical temperature $T_{\rm pc} = 155$~MeV~\cite{ALICE:2021rxa}. The extended data-to-model comparisons in Ref.~\cite{ALICE:2021rxa} further show the importance of recombination and radiative energy loss to describe the production of non-strange charm mesons.

\begin{figure}[tb!]
  \centerline{%
     \includegraphics[trim={0 0 9.43cm 0},clip,width=0.499\textwidth]{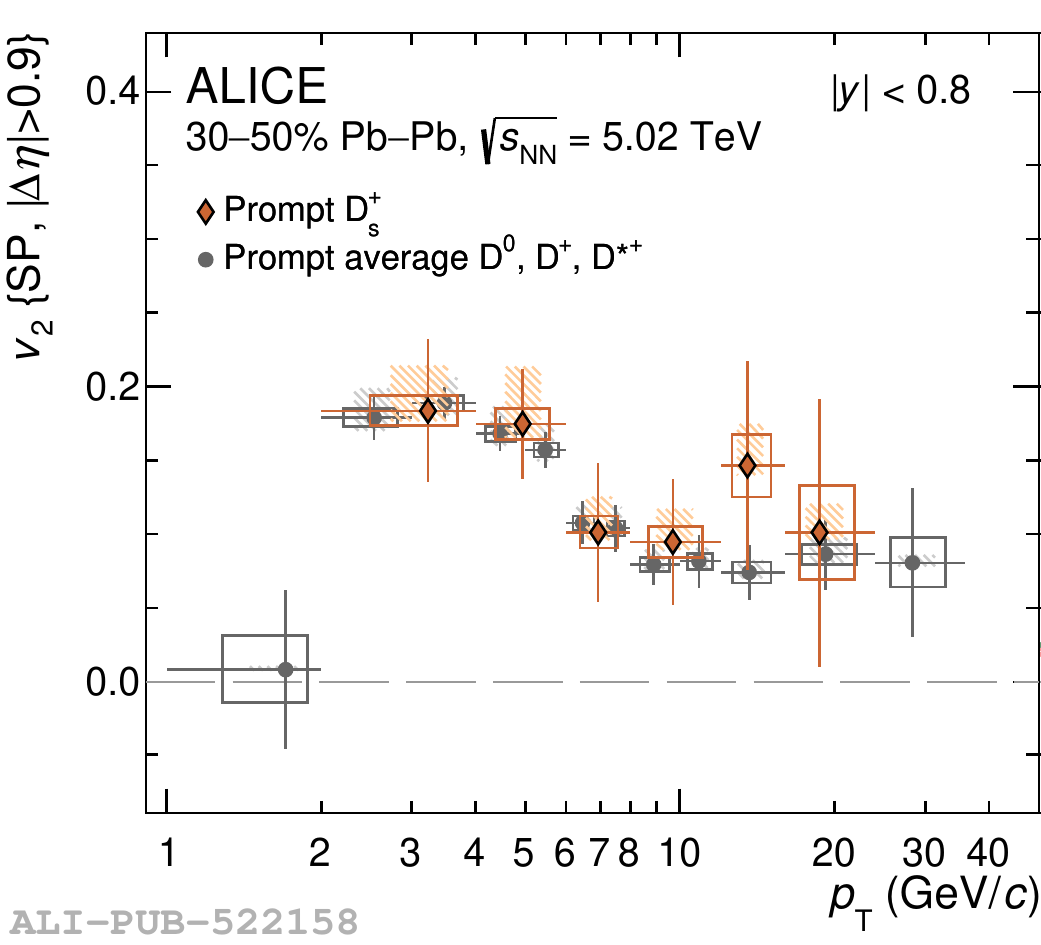}
  }
  \caption{The $v_2$ of prompt $\rm D_s^+$ in the 30--50\% centrality interval~\cite{ALICE:2021kfc} compared with that of non-strange $\rm D$ mesons~\cite{ALICE:2020iug}. The systematic uncertainty due to the feed-down subtraction is shown separately as shaded boxes.}
  \label{fig:Dsv2}
\end{figure}

Figure~\ref{fig:Dsv2} shows the elliptic flow for the strange $\rm D_s^+$ meson compared to the one of the non-strange $\rm D$ mesons in mid-central Pb--Pb collisions. The measured $\rm D_s^+$-meson $v_2$ is positive in the $2 < p_{\rm T} < 8$~GeV$/c$ interval with a significance of $6.4\sigma$~\cite{ALICE:2021kfc}. As argued in Ref.~\cite{He:2012df}, the comparison of the $v_2$ between the strange and non-strange D mesons provide sensitivity to the transport properties of the hadronic phase, since the $\rm D_s^+$ meson is expected to decouple earlier from the hadron gas due to its strange-quark content. Within the current uncertainties, it is, however, not possible to conclude about such a possible difference. The $\rm D_s^+$ nuclear modification factor and strange-to-non-strange $\rm D_s^+ / D^0$ yield ratio are presented as well in Ref.~\cite{ALICE:2021kfc}, indicating the importance of charm-quark hadronisation via recombination to ``pick up'' the enhanced abundance of strange quarks in the QGP medium.

\begin{figure}[tb!]
  \centerline{%
    \includegraphics[width=0.5\textwidth]{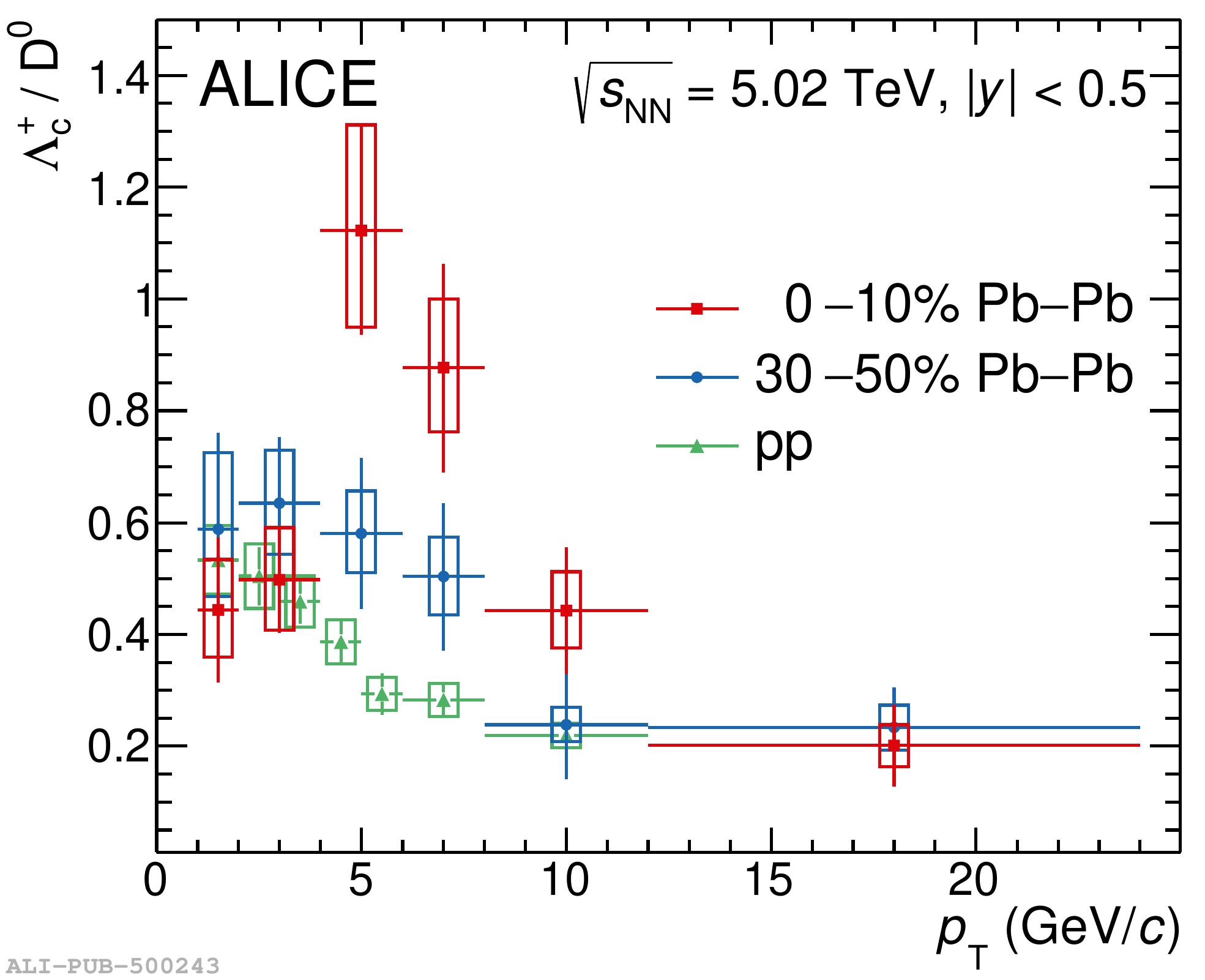}%
    \hfill
    \includegraphics[width=0.48\textwidth]{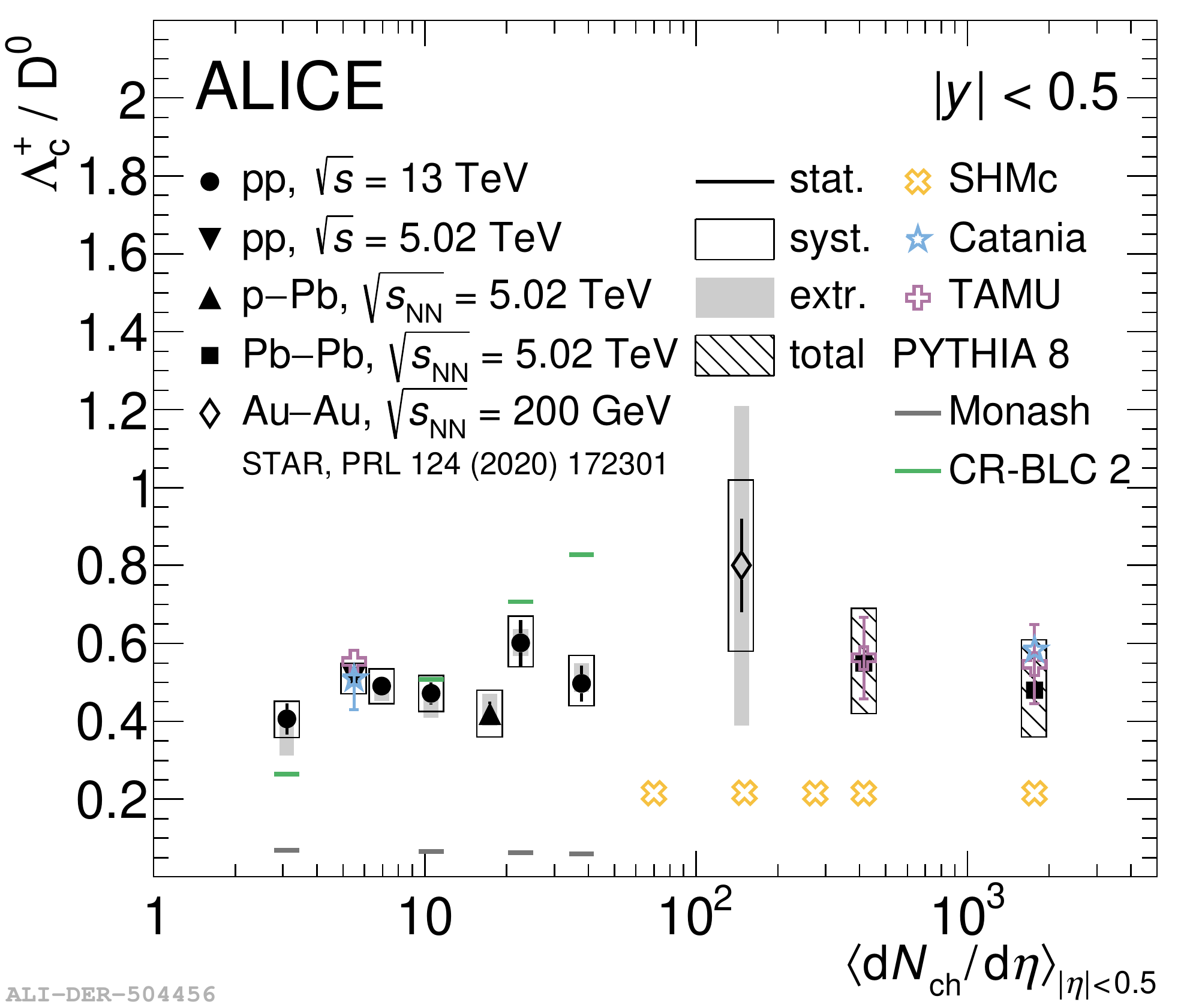}
  }
  \caption{Left: the $\rm \Lambda_c^+ / D^0$ yield ratio in central and mid-central Pb--Pb collisions~\cite{ALICE:2021bib} compared with the result from pp collisions~\cite{ALICE:2020wfu}. Right: The $p_{\rm T}$-integrated $\rm \Lambda_c^+ / D^0$ ratios as function of multiplicity in pp, p--Pb, Au--Au, and Pb--Pb collisions~\cite{ALICE:2021bib,ALICE:2020wfu,STAR:2019ank} compared with theoretical predictions~\cite{LcD0ptIntModels}.}
  \label{fig:LcD0}
\end{figure}

In the left panel of Fig.~\ref{fig:LcD0}, the $p_{\rm T}$-differential $\rm \Lambda_c^+ / D^0$ baryon-to-meson yield ratios in central and mid-central Pb--Pb collisions are shown~\cite{ALICE:2021bib}. Compared to the same ratio in pp collisions, which showed already a surprising enhancement with respect to the same ratio in $\rm e^+e^-$ collisions~\cite{ALICE:2020wfu}, the mid-central and central Pb--Pb ratios are further enhanced by $2.0\sigma$ and $3.7\sigma$ in the $4 < p_{\rm T} < 8$~GeV$/c$ interval, respectively. This, and the theory comparisons shown in Ref.~\cite{ALICE:2021bib}, indicate once more the importance of hadronisation via recombination for the description of charm-hadron production. To investigate if this enhancement is an overall enhancement of $\rm \Lambda_c^+$ production relative to the $\rm D^0$ one, as proposed by recombination models including light diquark states~\cite{LcD0ptIntenhanced}, the $\rm \Lambda_c^+$ yields are extrapolated to $p_{\rm T} = 0$. The corresponding $p_{\rm T}$-integrated $\rm \Lambda_c^+ / D^0$ ratios as function of multiplicity are shown in the right panel of Fig.~\ref{fig:LcD0}. The ratio values for Pb--Pb collisions are compatible with the ones at pp and p--Pb multiplicities~\cite{ALICE:2020wfu} within one standard deviation of the combined uncertainties, disfavouring the models expecting an overall enhancement of baryon production~\cite{LcD0ptIntenhanced}. The measured enhancement at intermediate $p_{\rm T}$ may instead be caused by altered $p_{\rm T}$ distributions of baryons and mesons due to the quark's phase-space distribution.

\section{Conclusion}

The ALICE Collaboration performed precision measurements of charm-hadron production in Pb--Pb collisions with the Run~2 data sample. The non-strange $\rm D$-meson yield was measured for the first time down to $p_{\rm T}=0$, and the precision and $p_{\rm T}$ reach of the charm-strange meson and charm baryon results was significantly improved with respect to previous measurements. The results point to charm-quark interactions with the medium constituents via collisional and radiative processes, indicate that low-$p_{\rm T}$ charm quarks thermalise with the medium and thus participate in the collective motion, and show the importance of the recombination process to describe charm-quark hadronisation. The upgraded ALICE detector for the LHC Runs 3 and 4 will allow for even more precise measurements and stronger constraints on model calculations thanks to the improved precision of the upgraded detectors and the larger data samples~\cite{Citron:2018lsq}.

\end{document}